\documentclass{jfm}

\usepackage{graphicx}
\usepackage{newtxtext}
\usepackage{newtxmath}
\usepackage{natbib}
\usepackage{siunitx}
\usepackage{hyperref}
\hypersetup{
    colorlinks = true,
    urlcolor   = blue,
    citecolor  = black,
}

\newcommand{\RomanNumeralCaps}[1]
\linenumbers

\title{Basal layer of granular flow down smooth and rough inclines: kinematics, slip laws and rheology}

\author{Teng Wang\aff{1,2},
Lu Jing\aff{1}\corresp{\email{lujing@sz.tsinghua.edu.cn}},
  Fiona C.Y. Kwok\aff{2},
  Yuri D. Sobral\aff{3},
  Thomas Weinhart\aff{4}
 \and Anthony R. Thornton\aff{4,5}}

\affiliation{\aff{1}Institute for Ocean Engineering, Shenzhen International Graduate School, Tsinghua University, Shenzhen, 518055, China 
\aff{2}Department of Civil Engineering, The University of Hong Kong, Hong Kong, China
\aff{3}Departamento de Matemática, Universidade de Brasília, Campus Universitário Darcy Ribeiro, 70910-900 Brasília, DF, Brazil 
\aff{4}Department of Thermal and Fluid Engineering, University of Twente, Enschede, 7500 AE, The Netherlands
\aff{5}Department of Applied Mathematics, University of Manchester, Manchester, M13 9PL, UK}

\begin{document}
\maketitle

\begin{abstract}

Granular flow down an inclined plane is ubiquitous in geophysical and industrial applications. On rough inclines, the flow exhibits Bagnold’s velocity profile and follows the so-called $\mu(I)$ local rheology. On insufficiently rough or smooth inclines, however, velocity slip occurs at the bottom and a basal layer with strong agitation emerges below the bulk, which is not predicted by the local rheology. Here, we use discrete element method simulations to study detailed dynamics of the basal layer in granular flows down both smooth and rough inclines. We control the roughness via a dimensionless parameter, $R_a$, varied systematically from 0 (flat, frictional plane) to near 1 (very rough plane). Three flow regimes are identified: a slip regime ($R_a \lesssim 0.45$) where a dilated basal layer appears, a no-slip regime ($R_a \gtrsim 0.6$) and an intermediate transition regime. In the slip regime, the kinematics profiles (velocity, shear rate and granular temperature) of the basal layer strongly deviate from Bagnold’s profiles. General basal slip laws are developed which express the slip velocity as a function of the local shear rate (or granular temperature), base roughness and slope angle. Moreover, the basal layer thickness is insensitive to flow conditions but depends somewhat on the inter-particle coefficient of restitution. Finally, we show that the rheological properties of the basal layer do not follow the $\mu(I)$ rheology, but are captured by Bagnold's stress scaling and an extended kinetic theory for granular flows. Our findings can help develop more predictive granular flow models in the future.

\end{abstract}



\section{Introduction}

Gravity-driven granular flows over an inclined plane has significant implications for geophysical and industrial applications, such as rock avalanches, debris flows, and transport of bulk materials~\citep{goldhirsch_rapid_2003, forterre_flows_2008, kamrin_advances_2024}. Although recent decades have seen many advances in the rheology~\citep{Jop2006,Kim_Kamrin2020,gaume_microscopic_2020} and flow rule of granular materials~\citep{Pouliquen1999,borzsonyi_flow_2007,wu_unified_2025}, a less tackled issue is the behaviour of granular flows near the boundary~\citep{delannay_granular_2017}. Among common boundary types of granular flows, i.e., flat, bumpy and erodible~\citep{delannay_towards_2007,richard_influence_2020}, as well as dissipative ~\citep{louge_granular_2015,malloggi_nonlocal_2015}, even the seemingly simplest non-erodible (smooth to rough) boundaries can lead to a variety of granular flow phenomena that escape a full understanding and, hence, accurate modelling \citep{Artoni2012,breard_basal_2020}. 

On sufficiently rough inclines (i.e., no slip occurs at the base), steady-state granular flows typically exhibit Bagnold's velocity profile in the bulk due to force balance and a rheology in which the shear stress scales with the square of the shear rate~\citep{Silbert2001}. A thin basal layer below the bulk has also been noted, where the flow is highly affected by the boundary~\citep{Louge2003, kumaran_dense_2008}. This basal layer can dominate in very thin flows (less than ten particles in thickness), leading to non-Bagnold flow velocity profiles~\citep{silbert_granular_2003,kamrin_nonlocal_2015}. In contrast to this basic picture on rough inclines, flows on smooth bases (a flat, frictional plane or an insufficiently rough plane) can display various distinct features, including basal slip~\citep{silbert_boundary_2002,Artoni2012,Jing2016}, plug flow~\citep{kumaran_effect_2013}, kinetic energy oscillation~\citep{silbert_boundary_2002}, ordering~\citep{weinhart_closure_2012,kumaran_transition_2012} and density inversion or the so-called conductive boundary layer~\citep{Taberlet2007,Berzi2024}. In particular, the rich dynamics within the dilute, highly agitated basal boundary layer can play a significant role in setting the effective slip velocity of granular flows~\citep{hui_boundary_1984,richman_boundary_1988,jenkins_flux_1997,jenkins_dense_2007,kumaran_dense_2008,tsang_granular_2019}, but we still lack a general slip law for granular flows on smooth and rough inclines~\citep{Artoni2012,gollin_extended_2017,breard_basal_2020,pol_unified_2023}. This drawback has largely hindered our modelling capability of natural hazards related to the high flow speed of landslides and granular avalanches.

To close this gap, we use discrete element method (DEM) simulations to study detailed dynamics of the basal boundary layer in granular flows on a series of smooth and rough inclined planes. The base roughness is varied systematically by changing the size and spatial distribution of fixed particles at the base. A threshold roughness is found, below which a basal layer appears regardless of flow and particle parameters. The kinematics and rheological data in the basal layer are analysed in several model frameworks, and a general basal slip law is developed for varying roughness, slope and flow thickness. Our results provide new insights into the dynamics of granular flows down an arbitrarily rough plane and may be used as benchmark cases for future development of granular flow theories.

\section{Methods}\label{sec:method}

\subsection {Simulation setup}

We use DEM to simulate steady, fully-developed granular flows down smooth and rough inclines. Figure~\ref{fig:basesetup} illustrates the simulation setup, which consists of a granular assembly with a constant length ($L/d=20$) and width ($W/d=10$), but varying thicknesses ($H/d=20$, $40$ and $60$), angles of inclination ($\theta=23\sim27^\circ$, increased at an increment of $1^\circ$) and base roughness ($R_a=0\sim0.85$, defined in \S\ref{sec:method-Ra}). The initial thickness $H$ is controlled by inserting an estimated number of particles and deleting particles above the designated $H$. The inclination is achieved by tilting the gravitational vector $(g\sin\theta, 0, -g\cos\theta)$ in the $xz$-plane, where $g=9.81$~\si{m/s^2}. Periodic boundaries are imposed in both the length ($x$) and width ($y$) directions. The top surface is free of constraint, while the bottom plane is roughened by a layer of fixed particles (detailed below). Particle interactions are modelled by the Hertzian contact model with a timestep $\Delta{t}=10^{-5}$ s and normal coefficient of restitution $e_n=0.8$ \citep{antypov_analytical_2011}. Coulomb’s friction criterion \citep{tsuji_lagrangian_1992} is applied with a coefficient of friction $\mu_p=0.5$. Other material properties include Young’s modulus $E=50$ MPa, Poisson’s ratio $\nu=0.4$, mean particle diameter $d=5$ mm and density $\rho_p=2650$ kg/m\textsuperscript{3}. Note that the particle diameter has a small distribution ($\pm0.1d$) to avoid internal crystalline structures. In addition, simulations with a controlled $H$ but a different particle size ($d = 10$ mm) are conducted to study particle size effects. We also vary $e_n$ from $0.1$ to $0.9$ for the flow particles (keeping $e_n=0.8$ for the base particles) to investigate how energy dissipation affects the basal layer thickness. Trial simulations are conducted twice with different random seeds to confirm repeatability.


\subsection {Generation of smooth and rough bases} \label{sec:method-Ra}
\begin{figure}
  \centerline{\includegraphics[width=1\textwidth]{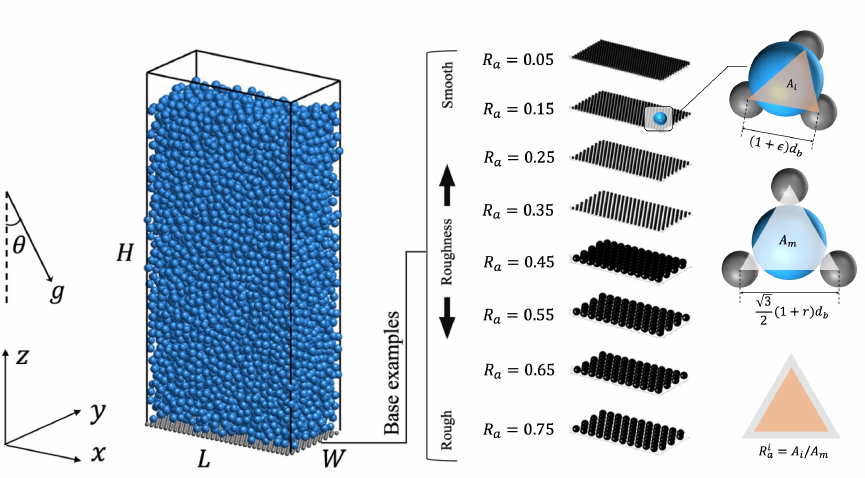}}
  \caption{Simulation setup, various roughness-controlled bottom surfaces and the definition of base roughness, $R_a$, at a local triangular void.}
\label{fig:basesetup}
\end{figure}

To control the base roughness for all granular flow simulations, we consistently generate the bottom plane with a layer of fixed, identical spherical particles of diameter $d_b$ (figure~\ref{fig:basesetup}). The fixed particles are centred at $z=-d_b/2$ (such that $z=0$ is the apparent bottom surface) and arranged on a uniform triangular lattice or in a random pattern, encompassing common strategies for rough base generation in granular flow simulations and experiments \citep{Goujon2003,weinhart_closure_2012,Jing2016}.

A roughness parameter must be unambiguously defined prior to further discussions, as both the flow-to-base particle size ratio ($r = d/d_b$) and the mean spacing ($\epsilon$) between base particles can affect the effective roughness condition. We characterise the base roughness using $R_a$ (meaning roughness by area), first proposed by \citet{Jing2016} to capture the effects of both $r$ and $\epsilon$, which is a dimensionless parameter varying from 0 (flat frictional plane) to 1 (extremely rough). As illustrated in figure~\ref{fig:basesetup}, the fixed particles on the bottom plane are first discretized into triangular regions using Delaunay triangulation. Then, the area of each triangular region ($A_i = (\sqrt{3}/4)(1+\epsilon)^2 d_b^2$, assuming equilateral triangles) is compared to a threshold area ($A_m$), which represents an idealized maximum-roughness scenario where a sphere of diameter $d_b$ can be trapped exactly by the triangular space (figure~\ref{fig:basesetup}b). It is easy to show that $A_m = (3\sqrt{3}/16)(1+r)^2 d_b^2$. As such, the local roughness at the $i$th triangular region is defined as $R_a^i = A_i/A_m$ and the overall base roughness $R_a$ is simply the mean of $R_a^i$ over all triangular regions of the entire base. For base particles generated on a uniform triangular lattice, all triangular regions are exactly equilateral (with side length of $(1 + \epsilon)d_b$) and we have
\begin{equation}
  R_a = \frac{4}{3}\left(\frac{1+\epsilon}{1+r}\right)^2.
  \label{eq:RaCal}
\end{equation}
The definition of $R_a$ also applies for randomly arranged base particles, where local variations in $R_a$ exist because the discretized triangular regions are not strictly equilateral (i.e., variations in $\epsilon$ for each $A_i$) and they have different areas (i.e., variations in $A_i$). In this case, we use the standard deviation of $A_i$ to estimate the variations of $R_a$; examples can be found in Table~\ref{tab:configuration} and Appendix~\ref{app:base}.
Moreover, alternative measures of the roughness, such as the roughness angle or penetration angle, $\sin\psi=(1+\epsilon)/(1+r)$, have been used in the literature~\citep{dippel_collision-induced_1996, gollin_extended_2017, berzi_shearing_2017,breard_basal_2020}. Note that our roughness parameter $R_a$ reduces precisely to $\sin\psi$ in two-dimensional configurations~\citep{Jing2016}. In the subsequent analysis, we make use of the relation $R_a=(4/3)\sin^2\psi$ to reinterpret the results of \citet{gollin_extended_2017} and \citet{breard_basal_2020}.

For a given $r$, it is required that $\epsilon \leq (\sqrt{3}/2)(1+r) - 1$, such that $R_a \leq 1$. Physically, this condition indicates that, if the triangular voids formed by the base particles are too large, flowing particles can percolate through or fill the voids, leading to a so-called void-filling mechanism and thereby reducing the apparent base roughness \citep{Goujon2003,Jing2016}. Here, to avoid such complications for $R_a \rightarrow 1$, we carefully choose $r$ and $\epsilon$ to restrict the range of $R_a$ from $0$ to $0.85$, where $R_a = 0$ means a flat base without base particles. We also conduct simulations with the same $R_a$ but different combinations of $r$ and $\epsilon$ (including regular and random arrangements of base particles) to demonstrate the relevance of using $R_a$ as a single indicator of the geometric roughness (see Appendix~\ref{app:base}). Detailed base configurations for various simulation series are listed in Table~\ref{tab:configuration}.

\begin{table}
  \begin{center}
\def~{\hphantom{0}}
  \begin{tabular}{lcccc}
      Simulation series &  Roughness ($R_a$)&Size ratio $(r)$&   Spacing ($\epsilon$)& Particle size ($d$)\\[3pt]
       Main (varying $H$ and $\theta$) &  0&-
& - & 0.005
\\
       &  0.05&5
& 0.162& 0.005
\\
       &  0.10&5
& 0.643& 0.005
\\
       &  0.15&2
& 0.006& 0.005
\\
       &  0.20&2
& 0.162& 0.005
\\
 &  0.25&2
& 0.299&0.005
\\
 & 0.30& 2
& 0.423&0.005
\\
 & 0.35& 2
& 0.537&0.005
\\
 & 0.40& 2
& 0.643&0.005
\\
 & 0.45& 2
& 0.743&0.005
\\
 & 0.47& 2
& 0.781&0.005
\\
 & 0.49& 2
& 0.819&0.005
\\
 & 0.51& 2
& 0.855&0.005
\\
 & 0.53& 2
& 0.891&0.005
\\
 & 0.55& 2& 0.927&0.005\\
 & $0.55^a$& 0.67& 0.070&0.005
\\
 & 0.57& 2
& 0.961&0.005
\\
 &  0.59&2
& 0.996&0.005
\\
 &  0.61&2
& 1.029&0.005
\\
 &  0.63&2
& 1.062&0.005
\\
 & 0.65& 2& 1.095&0.005\\
 &  $0.65^a$&0.67& 0.164&0.005
\\
 &  0.67&2
& 1.127&0.005
\\
 &  0.69&2
& 1.158&0.005
\\
 &  0.71&2
& 1.189&0.005
\\
 &  0.73&2& 1.220&0.005
\\
 & 0.75& 2& 1.250&0.005\\
 &  $0.75^a$&0.67& 0.250&0.005\\
 Same $R_a$ different bases&  0.25&2.5
& 0.516&0.005
\\
 &  0.25&2
& 0.299&0.005
\\
 &  0.25&1.67& 0.155&0.005\\
 & $0.24\pm0.06^b$& 2& -&0.005\\
 &  0.45&2.5
& 1.033&0.005
\\
 &  0.45&2
& 0.743&0.005
\\
 &  0.45&1.67& 0.549&0.005\\
 & $0.45\pm0.19^b$& 2& -&0.005\\
 Same $H$ different $d$&  0.15&2
& 0.006&0.010\\
 &  0.25&2
& 0.299&0.010\\
 &  0.35&2
& 0.537&0.010\\
 &  0.45&2& 0.743&0.010\\
 Varying $e_n$& 0.05& 5& 0.162&0.005\\
 & 0.15& 2& 0.006&0.005\\
 & 0.25& 2& 0.006&0.005\\
 & 0.35& 2& 0.537&0.005\\
 & 0.45& 2& 0.743&0.005\\
 & 0.55& 2& 0.927&0.005\\
 & 0.65& 2& 1.095&0.005\\
 & 0.75& 2& 1.250&0.005
\end{tabular}

  {\raggedright Notes: $a.$ Simulations with relatively large base particles ($r<1$), which show slightly different basal-layer shear rate profiles in figure~\ref{fig:kinematics_Ra}(b). $b.$ Simulations with randomly arranged base particles, where $R_a$ varies in a range quantified by the standard deviation of the void area. \par}

  \caption{Base roughness specifications for various simulations series.}
  \label{tab:configuration}
  \end{center}
\end{table}
 
\subsection {Steady flow conditions and mean flow fields}

To allow a granular flow to reach steady state, it is helpful to first determine the $h_{\text{stop}} (\theta)$ curve in the ($H$, $\theta$) parameter space, which is a curve delimiting no-flow (below the curve) and steady flow conditions \citep{Pouliquen1999}. The $h_{\text{stop}} (\theta)$ curve is usually measured using a sufficiently rough base, as lower base roughness can significantly affect $h_{\text{stop}} (\theta)$ \citep{weinhart_closure_2012}. In this study, the simulation protocol to determine $h_{\text{stop}} (\theta)$ is as follows. For a given $R_a$ and $H$, we initiate the granular flow at a certain slope angle and then decrease stepwise the slope angle at an increment  of $0.1^\circ$, holding for a sufficient period to check the flow state (accelerating, steady, or decelerating to stop). Such slope-reducing procedure continues until the system kinetic energy becomes negligible, at which the current slope angle is marked as $\theta_{\text{stop}}$ and the corresponding flow thickness (measured as twice the $z$-centre of mass for all particles) is marked as $h_{\text{stop}}$.

\begin{figure}
  \centerline{\includegraphics[width=1\textwidth]{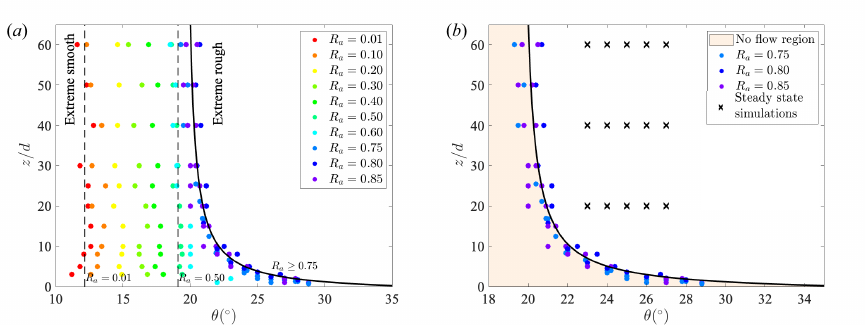}}
  \caption{(a) Results of the $h_\textrm{stop}$ curves for $R_a$ varying from $0.01$ to $0.85$. (b) The master $h_\textrm{stop}$ curve for sufficiently rough bases ($R_a>0.6$) and the main simulation conditions ($\theta,H/d$) for steady-state flows.}
\label{fig:hstop}
\end{figure}

Following this protocol, we conduct simulations to measure $h_{\text{stop}}(\theta)$ for $R_a$ varying from $0.01$ to $0.85$, as shown in figure~\ref{fig:hstop}(a). Interestingly, the $h_{\text{stop}}(\theta)$ curve is nearly vertical for small $R_a$, corresponding to a Coulomb friction-like condition \citep{Artoni2012}, and is shifted significantly to the right as $R_a$ is increased. For sufficiently rough bases ($R_a>0.6$), the $h_{\text{stop}}(\theta)$ curves converge to a master curve that is similar to those reported in previous studies \citep{Pouliquen1999,weinhart_closure_2012}. The master curve of $h_{\text{stop}}(\theta)$ is fitted to the following expression:
\begin{equation}
 \frac{h_\text{stop}}{d} = A\frac{\tan{\theta_2} - \tan{\theta}}{\tan{\theta} - \tan{\theta_1}},
  \label{eq:hstop}
\end{equation}
where $A=1.45$, $\theta_1=19.56^\circ$, and $\theta_2=37.24^\circ$ are fitting coefficients, which depend primarily on material properties as long as the base is sufficiently rough~\citep{Pouliquen1999,liu_effects_2024}. These fitting parameters are quantitatively similar to previous DEM simulations of spheres \citep{liu_effects_2024}. Finally, we show in figure~\ref{fig:hstop}(b) that our main simulations (marked by cross symbols) are chosen in the steady flow region well above the master curve of $h_{\text{stop}} (\theta)$.

Once a simulation reaches steady state (when the system kinetic energy does not vary in time), we compute various time-averaged mean flow fields based on the DEM data (following the formulations of \citet{Silbert2001}) in a series of $1d$-thick layers, including the volume fraction $\phi$, streamwise velocity $u$, shear rate $\dot{\gamma} = \partial u / \partial z$, granular temperature $T = (u_p - u)^2$, where $u_p$ is the instantaneous particle velocity, pressure (or normal stress) $P = \sigma_{zz}$, shear stress $\tau = \sigma_{xz}$, effective friction $\mu = \tau/P$ and inertial number $I = \dot{\gamma}d / \sqrt{P/\rho_p}$. All quantities are functions of the $z$-position, with $z=0.5d$ denoting the position of the lowermost data point.

\section{Flow kinematics}\label{sec:kinematics}

\subsection {Kinematics features of basal layer for varying base roughness} \label{sec:kinematics-basic}

\begin{figure}
  \centerline{\includegraphics{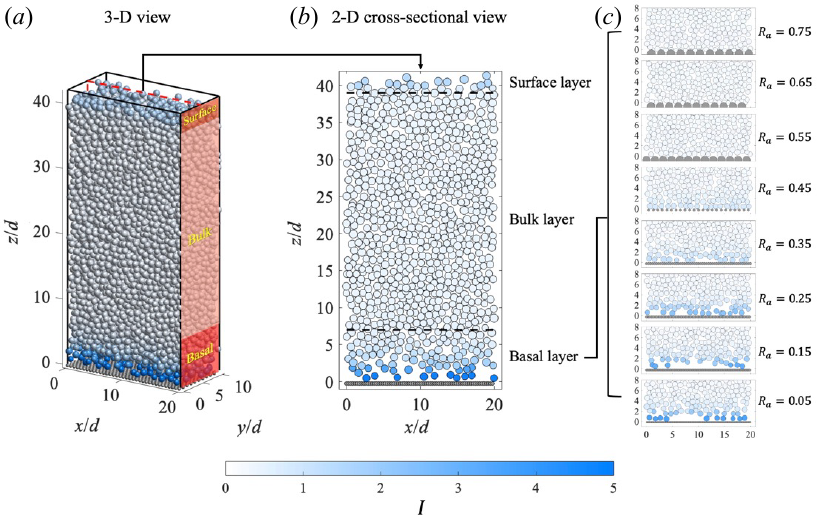}}
  \caption{Typical three-layer structure of inclined granular flows. (a) and (b) are an example case of $R_a=0.15$. (c) Basal layer for smooth ($R_a=0.05$) to rough ($R_a=0.75$) planes. All 2D snapshots are captured from 3D simulations at steady state, with particles coloured by the local inertial number $I$.}
\label{fig:snapshot}
\end{figure}

Figure~\ref{fig:snapshot} presents steady-state snapshots of the granular flow simulations for varying $R_a$ with a fixed set of flow conditions ($H/d=40$ and $\theta=25^\circ$), where the particles are coloured by layerwise values of the inertial number $I$. For an example case of $R_a=0.15$ (figures~\ref{fig:snapshot}a and b), which is relatively smooth, three layers of the flow can be identified: a dilated basal layer, a dense bulk layer (also known as the core), and a saltating surface layer. Although the three-layer structure is typical of steady-state flows down a bumpy base \citep{Louge2003}, it is clear from figure~\ref{fig:snapshot}(c) that $R_a$ controls the formation of the basal layer. A geometrically smoother plane leads to a basal layer that is much more dilated and energetic, with values of $I$ well above the typical value of $I<1$ for dense granular flows. Such basal layer behaviours have previously been noted as density inversion or the supported regime \citep{Taberlet2007,brodu_new_2015}, which can be understood as heating (i.e., increase of granular temperature) due to effective velocity slip of the basal layer particles along the bottom plane \citep{hui_boundary_1984,richman_boundary_1988,kumaran_dense_2008}.

\begin{figure}
  \centerline{\includegraphics[width=1\textwidth]{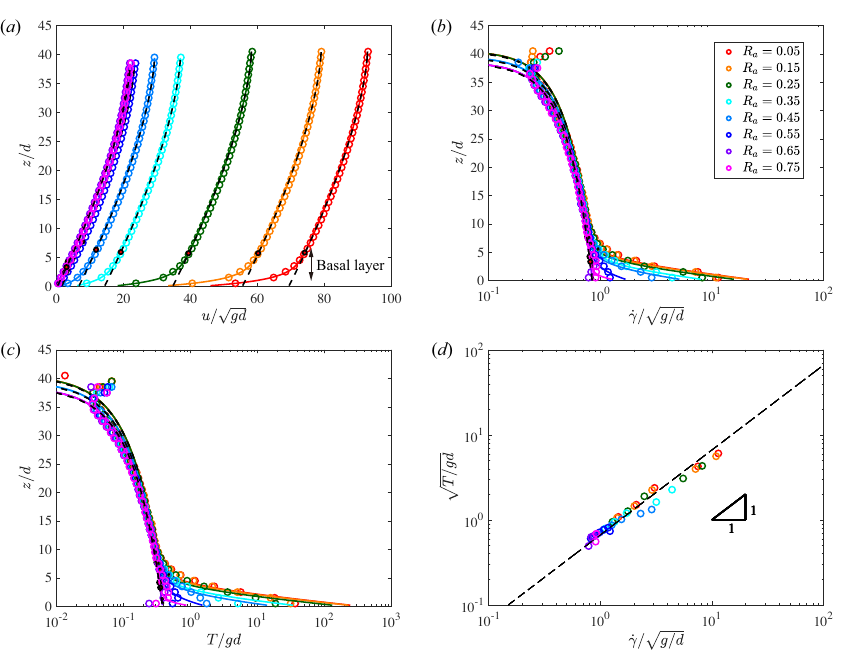}}
  \caption{Results for varying $R_a=0.05\sim0.75$ with fixed $H/d=40$ and $\theta=25^\circ$ in terms of (a) velocity, (b) shear rate, (c) granular temperature, and (d) the correlation between granular temperature and shear rate. The open circles denote the DEM results with their colour indicating $R_a$, while the dashed and solid curves in (a-c) are fittings based model (\ref{eq:uFit}) and (\ref{eq:uFitExp}), respectively.}
\label{fig:kinematics_Ra}
\end{figure}

To quantify the effects of $R_a$ on the kinematics features of the flow, we show flow-depth profiles of dimensionless velocity $u/\sqrt{gd}$, shear rate $\dot{\gamma}/\sqrt{g/d}$ and granular temperature $T/\sqrt{gd}$ in figure~\ref{fig:kinematics_Ra}, with $R_a$ varying from $0.05$ to $0.75$ at an increment  of $0.1$ ($H/d=40$ and $\theta=25^\circ$). Figure~\ref{fig:kinematics_Ra}(a) shows that a typical Bagnold velocity profile passing through the origin (no slip) is observed for $R_a\geqslant0.55$, but the profile is sequentially shifted to the right as $R_a$ becomes smaller (from blue to red symbols), indicating an increase in the basal slip velocity. The velocity profiles for all $R_a$ are parallel to Bagnold’s profile in the bulk, but a longer tail near the base is observed for lower-$R_a$ cases. Based on these observations, we use the following modified Bagnold velocity profile (dashed curves) to fit the simulation data (symbols),
\begin{equation}
 \frac{u_\textrm{Bag1}}{\sqrt{gd}} = \frac{u_{b1}}{\sqrt{gd}} + \frac{2}{3}\bar{I}\sqrt{\bar\phi\cos{\theta}} \left[ \left(\frac{h}{d}\right)^{3/2} - \left(\frac{h-z}{d}\right)^{3/2} \right],
  \label{eq:uFit}
\end{equation}
where $u_{b1}$ is a fitting parameter representing an effective slip velocity (see Appendix~\ref{app:Bagnold} for detailed derivation of the expression). Note that the flow thickness $h$ is defined by the $z$-coordinate of the uppermost data point and we use the bulk-averaged inertial number $\bar{I}$ and volume fraction $\bar\phi$ of each case for fitting. The basal layer thickness can be roughly considered as where the velocity data deviates from the dashed curve (manually marked by red filled circles), which seems to be insensitive to $R_a$ (see \S\ref{sec:thickness} for more discussions). Moreover, in Appendix~\ref{app:base} we present results of the same $R_a$ but different base particle arrangements, verifying that the use of $R_a$ as a single roughness indicator is reasonable.

The parallelness of the bulk velocity profiles in figure~\ref{fig:kinematics_Ra}(a) is more evident in figure~\ref{fig:kinematics_Ra}(b), where the shear rate profiles for all $R_a$ (symbols) overlap in the bulk region, following Bagnold’s shear rate profile (dashed curve), but deviate systematically in the basal layer. Specifically, a smaller $R_a$ corresponds to higher shear rates in the basal layer (from blue to red), reflecting the longer tails of the velocity profiles in figure~\ref{fig:kinematics_Ra}(a). Similar trends are found in figure~\ref{fig:kinematics_Ra}(c) for the granular temperature profiles, where the basal layer for low values of $R_a$ is significantly heated (and hence dilated as shown in figure~\ref{fig:snapshot}). Finally, we note that the similarity of the temperature and shear rate profiles is expected, because the following relation has been widely reported in steady granular flows \citep{daCruz2005, Kim_Kamrin2020, rognon_shear-induced_2021}:
\begin{equation}
\sqrt{T} = \alpha \dot{\gamma} d,
  \label{eq:Tgamma}
\end{equation}
where $\alpha = 0.72$ is a proportionality constant according to our simulation data in figure~\ref{fig:kinematics_Ra}(d). This correlation is used to derive the temperature profiles in figure~\ref{fig:kinematics_Ra}(c) (dashed curves); see also Appendix~\ref{app:Bagnold}. 

Now that the relations between velocity, shear rate and temperature are clear, it is possible to provide a more accurate functional form for the velocity profiles in figure~\ref{fig:kinematics_Ra}(a) and therefrom derive the shear rate (figure~\ref{fig:kinematics_Ra}b) and temperature profiles (figure~\ref{fig:kinematics_Ra}c). We propose a general expression for the velocity profile considering the detailed shape of the basal layer (solid curves in figure~\ref{fig:kinematics_Ra}a):
\begin{equation}
\frac{u_\textrm{Bag2}}{\sqrt{gd}} = \frac{u_{b2}}{\sqrt{gd}} +\frac{\Delta u}{\sqrt{gd}}\left[1-\exp \left(-\frac{z}{z_c}\right) \right] + \frac{2}{3}\bar{I}\sqrt{\bar\phi\cos{\theta}} \left[ \left(\frac{h}{d}\right)^{3/2} - \left(\frac{h-z}{d}\right)^{3/2} \right],
  \label{eq:uFitExp}
\end{equation}
where $u_{b2}$, $\Delta u$ and $z_c$ are fitting parameters with clear physical meanings. Specifically, $u_{b2} = u(z = 0)$ is the (fitted) true basal slip velocity, $\Delta u$ represents the difference between the true and effective slip velocities (i.e. $\Delta u=u_{b1}-u_{b2}$), and $z_c$ is a characteristic distance for the basal slip velocity to change towards the bulk. Note that the assumed exponential decay function makes it convenient to define the basal layer thickness $h_b = 3z_c$, corresponding to where the difference between the flow velocity and Bagnold’s profile drops to about 5\% (figure~\ref{fig:kinematics_Ra}a). This definition of $h_b$ is found to be robust for most simulation conditions in this work, which is used and discussed in \S\ref{sec:thickness}. Note also that equation (\ref{eq:uFitExp}) can be easily converted into equations for shear rate (derivative in $z$) and temperature profiles (using equation (\ref{eq:Tgamma})), where the exponential function captures the decay of slip-generated shear rates and temperatures in the basal layer, as shown in figures~\ref{fig:kinematics_Ra}b and \ref{fig:kinematics_Ra}c (solid curves). 

\subsection {Effects of slope angle and flow thickness}

\begin{figure}
  \centerline{\includegraphics[width=1\textwidth]{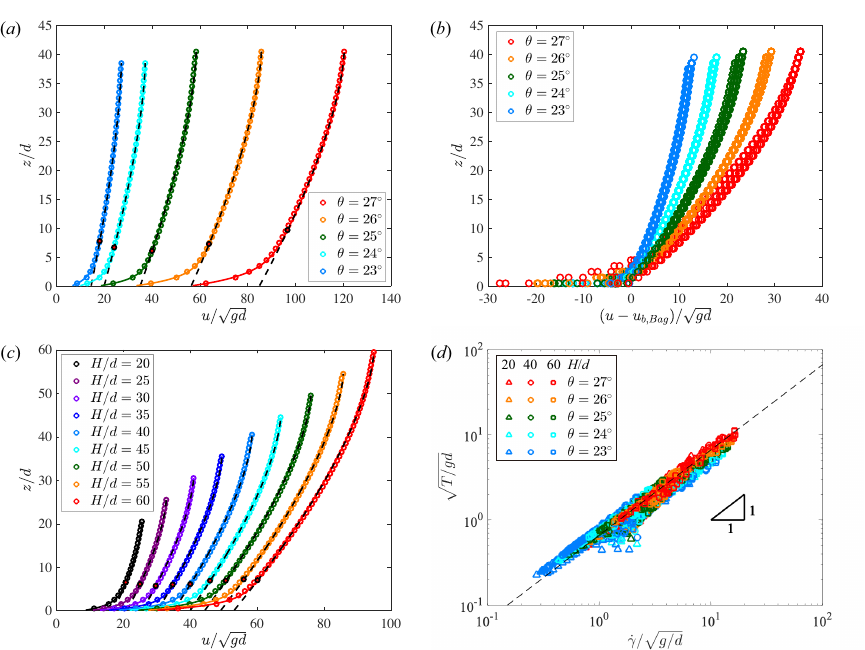}}
  \caption{(a) Velocity profiles for varying $\theta=23\sim28^\circ$ with fixed $H/d=40$ and $R_a=0.25$. (b) Shifted velocity profiles for varying $\theta$ and $R_a$ with fixed $H/d=40$. (c) Velocity profiles for varying $H/d=20\sim60$ with fixed $\theta=25^\circ$ and $R_a=0.25$. (d) Correlation between $T/gd$ and $\dot\gamma d$ in the basal layer for varying $R_a$, $H/d$ and $\theta$, which follows equation (\ref{eq:Tgamma}). The open circles in (a-c) denote DEM results, while the dashed and solid curves are fits to equations (\ref{eq:uFit}) and (\ref{eq:uFitExp}), respectively.}
\label{fig:kinematics_AH}
\end{figure}

To this point, we have investigated the basal effects with fixed slope angle $\theta$ and initial flow thickness $H$. Figure~\ref{fig:kinematics_AH}(a) shows the effects of varying $\theta$ from $23^\circ$ to $27^\circ$ with $H/d=40$ for a relatively smooth base ($R_a=0.25$). Both mean velocity and slip velocity increase significantly as $\theta$ is increased (blue to red data points). The velocity profiles can again be well described by (\ref{eq:uFit}) and (\ref{eq:uFitExp}), see black dashed and red solid curves, respectively, indicating that the bulk velocity follows Bagnold's scaling. The exponential tail in the basal layer seems to vary its shape as $\theta$ is varied, which is a rather non-trivial dependence; we address the detailed scaling of basal slip velocity in \S\ref{sec:slip}. Nevertheless, figure~\ref{fig:kinematics_AH}(b) shows that if one shifts the velocity profile to the origin, i.e., $u(z)-u_{b1}$, where $u_{b1}$ is obtained using (\ref{eq:uFit}), all velocity profiles for varying $R_a$ fall on branches defined only by $\theta$. In other words, the boundary effect decays rapidly within a few particles distance and has negligible influence beyond. Bulk particles can be viewed as a flow running atop the fast-moving basal layer, and the internal deformation rate within the bulk is determined by Bagnold's scaling or the local rheology (i.e., the shear stress scales with the square of the shear rate).  

In figure~\ref{fig:kinematics_AH}(c), we show velocity profiles for varying $H/d$ with a fixed slope ($\theta=25^\circ$) and smooth base ($R_a=0.25$). Similarly, the bulk flow follows Bagnold's scaling and the basal-layer velocity profile depends somewhat on $H/d$ (see \S\ref{sec:slip}). Intriguingly, the thickness of the basal layer, marked by red filled circles, does not grow with the overall flow thickness. This indicates that the basal layer is a local phenomenon determined by particle-scale interactions and that the excess energy (granular temperature) generated at the slippery bottom plane is dissipated via collisions in just a few particle layers. We address the scaling of the basal layer thickness in \S\ref{sec:thickness}.

Profiles of shear rate and granular temperature for varying $\theta$ and $H/d$ are omitted as they are similar to those in figure~\ref{fig:kinematics_Ra}. Rather, figure~\ref{fig:kinematics_AH}(d) shows the correlation $\sqrt{T} \propto \dot{\gamma}d$, consistent with (\ref{eq:Tgamma}), for all basal layer datasets with varying $R_a=0\sim0.55$, $\theta=23\sim27^\circ$ and $H/d=20\sim60$. This correlation further verifies that the velocity, shear rate and temperature profiles can be empirically described using (\ref{eq:uFitExp}), its $z$-derivative and (\ref{eq:Tgamma}) for all conditions studied here. Of course, it is also interesting to derive alternative theoretical descriptions of the kinematics profiles using kinetic theory of granular flow \citep{kumaran_dense_2008,Berzi2024}, as further discussed in \S\ref{sec:rheology}, but this is beyond the scope of this work. 

\subsection {Scaling of basal slip velocity: Navier slip formulation}\label{sec:slip}

\begin{figure}
  \centerline{\includegraphics[width=1\linewidth]{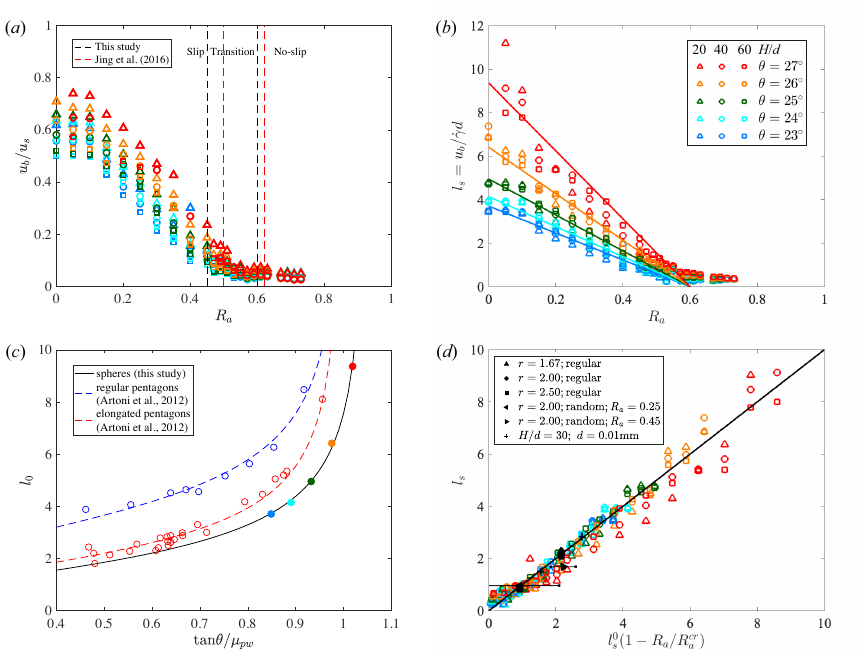}}
  \caption{Scaling of slip velocity. Main simulation results of (a) $u_b/u_s$ and (b) $l_s=u_b/\dot\gamma d$ vs.\ $R_a$ for various flow conditions ($\theta$ and $H/d$). (c) The flat-base slip length $l_s^0$ vs.\ $\tan\theta/\mu_{pw}$ from our simulations (filled symbols and solid curve) and previous work~\citep{Artoni2012} (open symbols and the corresponding curves). The curves are fits to (\ref{eq:sliplength0}). (d) All results of $l_s$ vs.\ $l_s^0(1-R_a/R_a^{cr})$ for varying $R_a$, $\theta$ and $H/d$. Coloured symbols represent the same cases as (a) and (b). The added symbols are additional simulations with different base configurations and particle sizes ($H/d=40$ and $\theta=23^\circ$).}
\label{fig:slip}
\end{figure}

Predicting the basal slip velocity is key to describing granular flows down an insufficiently rough incline. The empirical models (\ref{eq:uFit}) and (\ref{eq:uFitExp}) can only be used if the value of $u_{b1}$ or $u_{b2}$ is known. To investigate how the slip velocity depends on various simulation conditions ($R_a$, $\theta$ and $H/d$), we first plot $u_b/u_s$ vs.\ $R_a$ for $\theta=23\sim27^\circ$ and $H/d=20\sim60$ in figure~\ref{fig:slip}(a), where $u_b$ and $u_s$ are velocities measured at the lowermost (basal) and uppermost (surface) flow layers in a velocity profile, respectively, following \citet{Jing2016}. Despite the spread of the data, it is clear that $u_b/u_s$ decreases from $0.6\pm0.1$ towards zero, as $R_a$ is increased from $0$ to $0.6$, and remains zero (no-slip) beyond the threshold value of $R_a^{cr}=0.6$. Such slip-to-no-slip transition is consistent with previous DEM simulations in \citet{Jing2016}, as marked by vertical dashed lines in figure~\ref{fig:slip}(a).  

Although basal slip appears to be controlled by $R_a$, the data scatter in figure~\ref{fig:slip}(a) suggests that $\theta$ and $H/d$ can also affect the results. However, there is no straightforward way to express $u_b/u_s$ as a function of $\theta$ and $H/d$. In figure~\ref{fig:slip}(b), we rescale $u_b$ as $l_s=u_b/\dot\gamma_b d$ following a Navier slip velocity formulation proposed by \citet{Artoni2012}, where $l_s$ is called the slip length and $\dot\gamma_b$ is the basal shear rate measured in the lowermost flow layer. In this way, we effectively remove the $H/d$-dependence and find a clear monotonic dependence of $u_b/\dot\gamma_b d$ vs.\ $R_a$ on $\theta$ (from blue to red data, $\theta$ increases). For simplicity, we fit the data series for each $\theta$ in figure~\ref{fig:slip}(b) by a straight line passing through $(R_a^{cr},0)$, and the intersection of this line with the vertical axis is denoted as $l_s^0$, meaning the slip length for a flat base case ($R_a=0$). Moreover, to consider the effects of $\theta$, we exploit the fact that in steady-state inclined flows, $\mu=\tau/P=\tan\theta$ due to force balance \citep{Silbert2001}, where $\mu$ is the effective friction of the flow. As such, we propose the following semi-empirical scaling law for the slip velocity,
\begin{equation}\label{eq:sliplength}
\left\{ \begin{aligned} 
  l_s&:=\frac{u_b}{\dot\gamma_b d} = l_s^0 \left( 1-\frac{R_a}{R_a^{cr}}\right), R_a\lesssim R_a^{cr}\\
  u_b&=0, \rm{otherwise}
\end{aligned} \right.
\end{equation}
where $R_a^{cr}=0.6$ is an empirical threshold roughness below which slip occurs and $l_s^0$ is the flat-base slip length established by \citet{Artoni2012}, 
\begin{equation}
l_s^0:=\frac{u_b}{\dot\gamma_b d}\Big|_{R_a=0}=a\left( \frac{\mu/\mu_{pw}}{c-\mu/\mu_{pw}}\right)^b,
  \label{eq:sliplength0}
\end{equation}
where $a$, $b$ and $c$ are fitting parameters, and $\mu_{pw}$ is the coefficient of friction between particles and the bottom wall, which mainly affects the results when geometric roughness ($R_a$) is absent \citep{thornton_frictional_2012,Jing2016}; here, we fix $\mu_{pw}=0.5$. Fitting parameters $a$, $b$ and $c$ are material dependent and, as shown in figure~\ref{fig:slip}(c), we find $a=2.09$, $b=0.42$ and $c=1.05$ in our 3D simulations of spherical particles. Interestingly, these values are comparable to the 2D simulation results reported by \citet{Artoni2012}, i.e., $a=2.2$, $b=0.42$ and $c=1$ for elongated polygons and $a=3.66$\footnote{The reported value in \citet{Artoni2012} was $a=4.23$, which does not match the data well.}, $b=0.33$ and $c=1$ for regular polygons, as also included in figure~\ref{fig:slip}(c). Moreover, we find that flat-base simulations cannot reach steady state at slope angles close to the particle-wall contact friction angle ($\theta\rightarrow\tan^{-1}\mu_{pw}$), hence the missing data points for $(\theta=27^\circ, R_a=0)$ in figure~\ref{fig:slip}(b) and the diverging curves at $\tan\theta/\mu_{pw}\rightarrow1$ in figure~\ref{fig:slip}(c). 

In figure~\ref{fig:slip}(d), we plot $l_s$ vs.\ $l_s^0(1-R_a/R_a^{cr})$ according to the scaling law (\ref{eq:sliplength}) and (\ref{eq:sliplength0}), which collapses results for all simulation conditions ($R_a=0\sim0.75$, $\theta=23\sim27^\circ$ and $H/d=20\sim60$) onto a single curve. Moreover,  figure~\ref{fig:slip}(d) include two sets of additional simulations (black symbols): one with the same $R_a$ but different ($r$, $\epsilon$) to demonstrate the robustness of using $R_a$ as a unified roughness parameter (see also Appendix~\ref{app:base}), and another with the same $H$ but varying $d$ to confirm that the particle size is the correct length scale in $l_s$. Detailed simulation parameters for the added cases are listed in Table~\ref{tab:configuration}. Good agreement is found for all cases between the model predictions and simulation results, which validates our slip law as a generalisation of the flat-base slip law proposed by \citet{Artoni2012} accounting for the effects of geometric roughness. 


\subsection {Scaling of basal slip velocity: Temperature-based formulation}\label{sec:slip-T}

\begin{figure}
  \centerline{\includegraphics[width=1\linewidth]{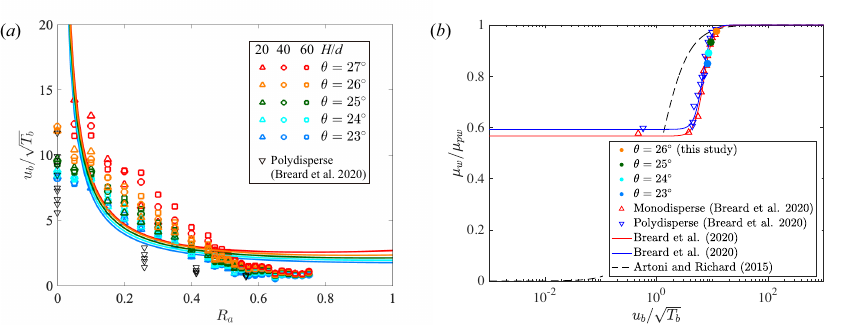}}
  \caption{Granular temperature-based scaling of slip velocity. (a) Rescaled slip velocity $u_b/\sqrt{T_b}$ as a function of $R_a$ for varying $H/d$ and $\theta$ (same simulations as in figure~\ref{fig:slip}. The curves are predictions of model (\ref{eq:slip-ktgf}); see text. The black triangles are polydisperse inclined flow simulation results of \citet{breard_basal_2020}. (b) Rescaled effective friction $\mu_w/\mu_{pw}$ as a function of $u_b/\sqrt{T_b}$ for flat-wall simulations ($R_a=0$), following two scaling laws from the literature \citep{artoni_effective_2015,breard_basal_2020}. Simulation data of \citet{breard_basal_2020} are also included.}
\label{fig:slip_T}
\end{figure}

An alternative slip law is to scale $u_b$ by the square root of the basal temperature, $\sqrt{T_b}$, as found in previous DEM simulations~\citep{artoni_effective_2015,breard_basal_2020} and derived using kinetic theory of granular flows~\citep{hui_boundary_1984, richman_boundary_1988, jenkins_dense_2007, berzi_shearing_2017, gollin_extended_2017}. Figure~\ref{fig:slip_T}(a) presents rescaled results of $u_b/\sqrt{T_b}$ vs.\ $R_a$ for all main simulations, which appears to collapse better than figure~\ref{fig:slip}(b) in terms of the $\theta$ dependence. Similar three-dimensional DEM simulations of spheres on flat- and rough inclined surfaces by \citet{breard_basal_2020} show similar results (although their results are systematically lower at medium values of $R_a$), as reproduced in figure~\ref{fig:slip_T}(a). Note that \citet{breard_basal_2020} used the roughness angle, $\sin\psi$, which we convert to $R_a$ using the relation $R_a=(4/3)\sin^2\psi$ (see \S\ref{sec:method}). Note also that \citet{breard_basal_2020} calculated the slip velocity by considering the mean velocity of all particles contacting the base, whereas we extract $u_b$ from the lowermost binning layer, which might be responsible for the differences in the results. Nevertheless, the agreement for $R_a=0$ is excellent, as shown in figure~\ref{fig:slip_T}(b) and discussed shortly.  

The relation between $u_b/\sqrt{T_b}$ and base roughness has been derived as a boundary condition in the granular kinetic theory \citep{richman_boundary_1988,gollin_extended_2017},
\begin{equation}
    \frac{u_b}{\sqrt{T_b}}=\left(\frac{\pi}{2}\right)^{1/2}\mu_b \left[\frac{1}{3\sqrt{2}J_b}\frac{2^{3/2}J_b-5F_b(1+B_b)\sin^2\psi}{2(1-\cos\psi)/\sin^2\psi-\cos\psi}+\frac{5F_b}{2^{1/2}J_b}\right]
  \label{eq:slip-ktgf}
\end{equation}
where $B=\pi[1+5/(8\phi g_0)]/(12\sqrt{2})$, $F=(1+e_n)/2+1/(4\phi g_0)$, $J$ and $g_0$ are given in Appendix~\ref{app:stress}, and the subscript $b$ denotes variables computed at the base (lowermost layer of our binning results). For simplicity, we take $\mu_b=\tan\theta$ and $\phi_b$ to be a constant for each $R_a$ and $\theta$, although both can vary in the basal layer (see \S\ref{sec:rheology}); we verify that considering the variations in $\phi_b$ does not change the model prediction significantly. The predictions of (\ref{eq:slip-ktgf}) for various $\theta$ are plotted in figure~\ref{fig:slip_T}(a), where we convert $\sin\psi$ and $\cos\psi$ to $R_a$ using $R_a=(4/3)\sin^2\psi$ and $\sin^2\psi+\cos^2\psi=1$. While the predictions do not precisely capture our DEM results for each $\theta$ (indicated by colours) and they diverge at $R_a\rightarrow0$, the overall agreement is decent. This motivates further improvement of the boundary condition for kinetic theory. 

Finally, for flat walls ($R_a=0)$, scaling relations between $u_b/\sqrt{T_b}$ and $\mu_w/\mu_{pw}$ have been established in inclined and wall-confined flows~\citep{artoni_effective_2015,richard_influence_2020,breard_basal_2020}, where $\mu_w$ is the effective friction measured at the lateral or bottom wall. Figure~\ref{fig:slip_T}(b) shows the comparison of our data and previous models. While the empirical model of \cite{artoni_effective_2015} (dashed curve) predicts smaller $u_b/\sqrt{T_b}$ in general for wall-confined flows (where creeping occurs easily), the inclined flow results of \cite{breard_basal_2020} agree with our results very well (red and blue for mono- and polydisperse flows, respectively). We follow the empirical model of \cite{breard_basal_2020} to fit the data,
\begin{equation}
    \mu_w=a+\frac{\mu_{pw}-a}{1+b(u_b/\sqrt{T_b})^c}     
  \label{boundary Breard}
\end{equation}
where $a$, $b$, and $c$ are empirical fitting coefficients. Specifically, for monodisperse flows, $a=0.283$, $b=27178$, and $c=5.22$, whereas for polydisperse flows, $a=0.296$, $b=55783$, and $c=5.71$. Note that our data are concentrated at the higher end of the curves, because the slope angles we explore are well above the angle of repose of the material (see \S\ref{sec:method}).



\subsection {Scaling of basal layer thickness}\label{sec:thickness}

\begin{figure}
  \centerline{\includegraphics[width=1\textwidth]{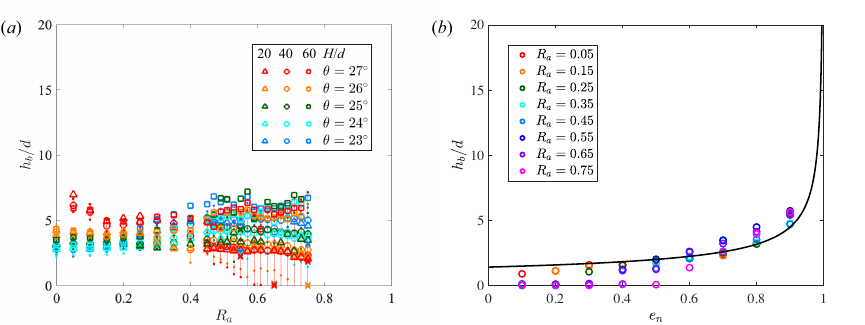}}
  \caption{Dependence of basal layer thickness $h_b/d$ on (a) the base roughness $R_a$ ($H/d=20, 40, 60$; $\theta=23\sim27^\circ$) and (b) the coefficient of restitution $e_n$ ($H/d=40$, $\theta=25^\circ$). Error bars in (a) indicate uncertainties of obtaining $h_b=3z_c$ via fitting the velocity profiles to model (\ref{eq:uFitExp}), where $z_c$ is a fitting parameter. The crosses in (a) represent the rough-base cases ($R_a=0.55$, $0.65$, and $0.75$) where the base particles are larger than the flow particles ($r=0.67$). The solid curve in (b) is a fit to $h_b/d=\beta/\sqrt{1-e_n}$ according to \citet{kumaran_dense_2008}; see text.}
\label{fig:thickness}
\end{figure}

An intriguing observation in figures~\ref{fig:kinematics_Ra} and \ref{fig:kinematics_AH} is that the basal layer thickness $h_b$ is generally insensitive to all simulation conditions ($R_a$, $\theta$, $H/d$). To investigate this, we plot $h_b/d$ vs.\ $R_a$ for varying $\theta=23\sim27^\circ$ and $H/d=20\sim60$ in figure~\ref{fig:thickness}(a). Despite the scatter, partly due to difficulties in consistently defining $h_b$ (we use $h_b=3z_c$ and $z_c$ is a fitting parameter; see \S\ref{sec:kinematics-basic}), it is evident that the basal layer thickness is only a few particle diameters even for significantly increased $H/d$. More specifically, for $R_a \lesssim 0.45$ (slip regime), $h_b\approx3d$ is nearly a constant except for the $\theta=27^\circ$ cases (red symbols). For the transitional ($0.45 \lesssim R_a \lesssim 0.6$) and no-slip regimes ($R_a \gtrsim 0.6$), the scatter of the data increases, likely because $h_b=3z_c$ depends somewhat on the goodness of fitting (error bars) when expression (\ref{eq:uFitExp}) is used to fit the velocity profiles (figures~\ref{fig:kinematics_Ra} and \ref{fig:kinematics_AH}). For rough bases ($R_a \gtrsim 0.6$), the assumption of an exponential tail in expression (\ref{eq:uFitExp}) may not hold, especially for $r<1$ cases where the shear rate in the basal layer points to zero due to the geometric constrains of the roughness (see figure~\ref{fig:kinematics_Ra}b), hence greater uncertainties in the results (crosses in figure~\ref{fig:thickness}a). Nevertheless, it is interesting that a basal layer, where the velocity deviates from Bagnold's, occurs even for the no-slip regime and the basal layer thickness is not wildly different from the slip regime. The systematic $\theta$-dependence of $h_b$ in the no-slip regime ($R_a \gtrsim 0.6$) is somewhat consistent with a previous study \citep{thornton_multi-scale_2013}.

The observation that $h_b$ does not grow with $H$ indicates that the basal layer dynamics is mainly controlled by particle-scale mechanisms. In kinetic theory of granular flows, this can be understood as the balance of fluctuation kinetic energy among three terms: energy generation due to shear work, conduction along the temperature gradient and dissipation due to inelastic collisions. Following this idea, \citet{kumaran_dense_2008} derived an expression for a conduction length, $\delta=d/\sqrt{1-e_n}$, near the base of an inclined granular flows, which is the length over which the rate of energy conduction is comparable to the rate of collisional dissipation. Interestingly, $\delta/d$ does not depend on flow conditions, as observed here, but is dependent on the material property $e_n$. To confirm the relevance of this scaling, we run additional simulations with varying $e_n=0.1\sim0.9$ and $R_a=0.05\sim0.75$ for fixed $\theta=25^\circ$ and $H/d=40$. As shown in figure~\ref{fig:thickness}(b), the scaling $h_b/d=\beta/\sqrt{1-e_n}$ agrees reasonably with our data for the slip regime, $R_a\leqslant0.45$, where $\beta=1.42$ is a fitting parameter. The slight disagreement may be attributed to the simplification of the conduction length derivation where particle friction was not considered, as noted by \citet{kumaran_dense_2008}.

\section{Rheology}\label{sec:rheology}

\begin{figure}
  \centerline{\includegraphics[width=1\linewidth]{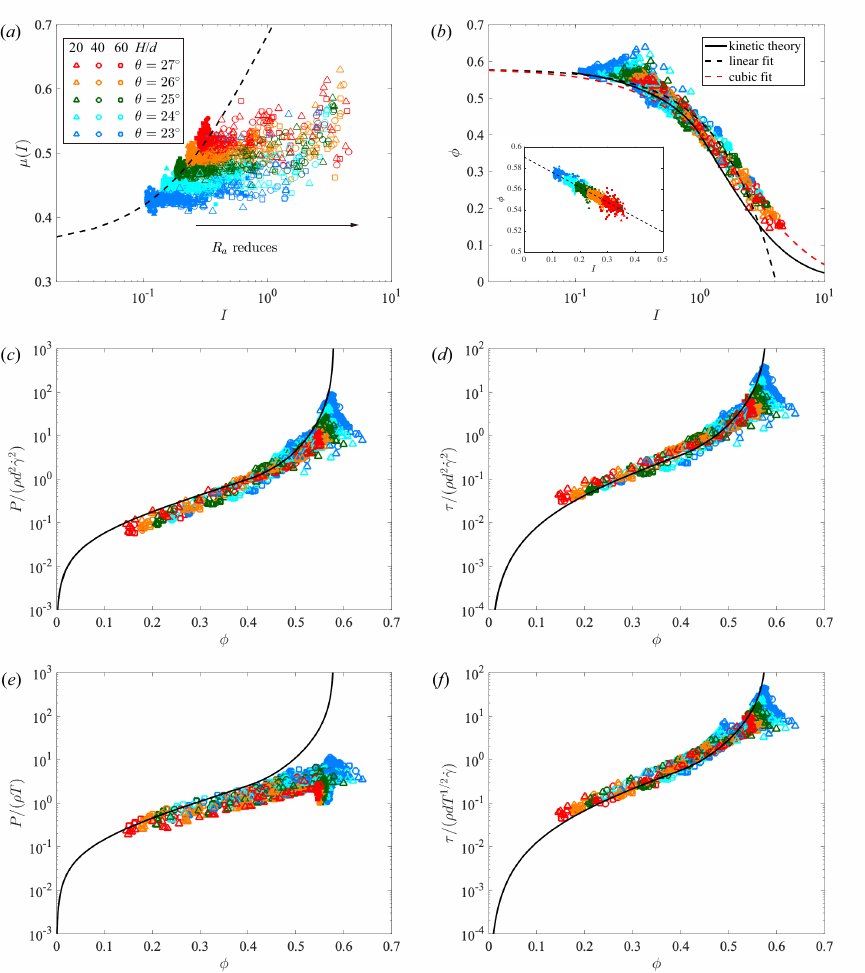}}
  \caption{Measurements (DEM simulations, solid and open symbols) and predictions from kinetic theory (lines) of pressure and shear stress in steady granular flows}
\label{fig:rheology}
\end{figure}

In the preceding section, we show that granular flows down an insufficiently rough plane ($R_a\lesssim0.6$) can exhibit a dilute, highly agitated basal layer beneath the dense bulk, where the kinematics profiles (velocity, shear rate, and granular temperature) deviate from Bagnold's profiles or from what the local $\mu(I)$ rheology can predict. Next, we explore friction (or stress) and volume fraction data in the basal and bulk layers to further understand the basal layer dynamics from a rheological point of view (figure~\ref{fig:rheology}). 

Figures~\ref{fig:rheology}(a) and (b) show $\mu(I)$ and $\phi(I)$ graphs to test the local rheology established in dense granular flow studies \citep{Jop2006,forterre_flows_2008}. Here, filled and open symbols are data from the bulk and basal layers, respectively, while symbol shapes and colours indicate the simulation conditions ($H/d$, $\theta$). Clearly, the bulk friction data (filled symbols in figure~\ref{fig:rheology}a) follow the standard local rheology (solid curve),
\begin{equation}
    \mu = \mu_s + \frac{\mu_{\infty} - \mu_s}{\left(I_0/I + 1\right)},
    \label{eq:muI}
\end{equation}
where $\mu_s=0.36$, $\mu_{\infty}=0.98$, and $I_0=0.90$ are fitting coefficients based on the bulk data. However, the data obtained from the basal layers (open symbols in figure~\ref{fig:rheology}a) deviate significantly from the standard $\mu(I)$ curve. For a given $\theta$ (a given colour), while the values of $\mu$ are more or less constant due to force balance, i.e., $\mu=\tan\theta$, the $I$ values can increase well beyond $1$, reflecting the high shear rates near the bottom plane due to basal slip (\S\ref{sec:kinematics}). In other words, the one-to-one relation between $\mu$ and $I$ (i.e., local rheology) breaks down in the basal layer.

Likewise, the volume fraction results from the bulk (filled symbols in figure~\ref{fig:rheology}b and inset) follow the standard local rheology (solid curve),
\begin{equation}
\phi = \phi_c - \kappa I,
  \label{eq:phiI_linear}
\end{equation}
where $\phi_c=0.58$ is the critical (maximum) volume fraction in the quasi-static limit ($I \rightarrow 0$) and $\kappa=0.14$ is a fitting parameter. Data from the basal layers (open symbols in figure~\ref{fig:rheology}b) deviate from this linear approximation, but they fall on a nonlinear master curve. That is, the low values of $\phi$ in the dilated basal layers can be roughly captured by the high values of $I$ (which encodes the effects of shear rate and pressure). It is possible to empirically model this $\phi$-$I$ relation using a cubic function (see also figure~\ref{fig:rheology}b inset),
\begin{equation}
\phi = \frac{\phi_c}{(\kappa'I+1)^3},
  \label{eq:phiI_nonlinear}
\end{equation}
where $\kappa'=0.13$ is a fitting parameter. Model (\ref{eq:phiI_nonlinear}) can be viewed as a generalization of the linear $\phi(I)$ model (\ref{eq:phiI_linear}) towards the dilute, high-inertial-number limit.


Since the local rheology (\ref{eq:muI}) does not hold in the basal layer, we seek alternative models for the granular stresses. Recently, \citet{Berzi2024} showed that Bagnold's rheology and the granular kinetic theory can both capture the stresses (pressure and shear stress) in terms of shear rate, temperature and packing fraction for a wide range of dilute to dense flow configurations, encompassing inertial numbers from $10^{-3}$ to $\sim 2$. In figures~\ref{fig:rheology}(c) and (d), we follow Bagnold's stress scaling to plot $P/(\rho_p d^2 \dot\gamma^2)$ and $\tau/(\rho_p d^2 \dot\gamma^2)$ as functions of $\phi$, respectively. All data points from both the bulk and basal layers collapse on a master curve in each plot. Moreover, these data are captured by the Bagnold-like stress formulation suggested by \citet{Berzi2024} (solid curves; see Appendix~\ref{app:stress} for derivation),
\begin{equation}
    p=\frac{2J\phi[1+2(1+e_n)\phi g_0]}{15(1-e_\textrm{eff}^2)}\left[1+\frac{26(1-e_\textrm{eff})}{15}\frac{\mathcal{H}(\phi-0.49)(\phi-0.49)}{0.64-\phi}\right]\rho_pd^2\dot{\gamma}^2,
    \label{eq:BagP}
\end{equation}
\begin{equation}
    \tau=\frac{8J\phi^2g_0}{5\pi^{1/2}}\left[\frac{2J}{15(1-e_\textrm{eff}^2)}\right]^{1/2}\left[1+\frac{26(1-e_\textrm{eff})}{15}\frac{\mathcal{H}(\phi-0.49)(\phi-0.49)}{0.64-\phi}\right]\rho_pd^2\dot{\gamma}^2,
    \label{eq:BagTau}
\end{equation}

\noindent where the expressions for the radial distribution function $g_0$ and dimensionless term $J$ are provided in Appendix~\ref{app:stress}, the effective coefficient of restitution $e_\textrm{eff}=e_n-(\pi/2)\mu_p+(9/2)\mu_p^2$ is derived for nearly elastic spheres and $\mathcal{H}(\cdot)$ is the Heaviside function. Note that the only parameters in (\ref{eq:BagP}) and (\ref{eq:BagTau}) are $e_n=0.8$ and $e_\textrm{eff}=0.63$ according to our model input (\S\ref{sec:method}).

Lastly, we plot $P/(\rho_pT)$ and $\tau/(\rho_p dT^{1/2}\dot{\gamma})$ as functions of $\phi$, respectively, in figures~\ref{fig:rheology}(e) and (f), following the kinetic theory stress models suggested by \cite{Berzi2024}; see Appendix~\ref{app:stress}. Again, all data points fall onto a single master curve in each plot, indicating the potential applicability of kinetic theory in describing the basal layer dynamics. The detailed dependence of these dimensionless stresses on the packing fraction (solid curves), derived for granular gases of frictional particles \citep{Berzi2024}, is given by
\begin{equation}
    p=\phi[1+2(1+e_n)\phi g_0]\rho_pT,
    \label{eq:KTGF-P}
\end{equation}
\begin{equation}
    \tau=\frac{8J\phi^2g_0}{5\pi^{1/2}}\rho_p d T^{1/2}\dot{\gamma}.
    \label{eq:KTGF-tau}
\end{equation}

Models (\ref{eq:KTGF-P}) and (\ref{eq:KTGF-tau}) capture the main feature of the data in figures~\ref{fig:rheology}(e) and (f), but discrepancies are observed between the predictions (solid curves) and DEM results. This mismatch may be attributed to kinetic theory assumptions such as the single-particle velocity distribution function \citep{Berzi2024}. Nevertheless, the overall success of these kinetic theory models reinforces that the basal layer dynamics is controlled by the fluctuation kinetic energy, which is generated by shear work due to slip at the base and dissipated primarily due to inelastic collisions. Future work is expected to extend kinetic theory models \citep{richman_boundary_1988, kumaran_dense_2008, gollin_extended_2017} to predict the entire kinematics and rheological profiles in the basal layer as a boundary-value problem for all flow conditions explored here.

\section{Conclusions}
\label{sec:conlusions}

It has been long noticed that granular flows down an inclined plane can develop a dilute, highly agitated basal layer beneath the bulk, where the flow kinematics deviate from standard Bagnold's profiles and where local rheology breaks down. However, a full picture of the basal layer dynamics remains unclear. Here, we use extensive DEM simulations of granular flows down smooth to rough inclines, with various flow conditions, to provide insights into the formation, kinematics, rheology and the associated scaling laws of the basal layer. 

We find that the formation of a dilated basal layer is primarily controlled by the base roughness. As the roughness $R_a$ is varied from zero (flat base) to $0.75$ (sufficiently rough), three regimes are identified: a slip regime for smooth bases ($R_a\lesssim0.45$) where a dilated basal layer of several particle diameters appears, an intermediate regime ($0.45 \lesssim R_a \lesssim 0.6$) where slip-to-no-slip transition occurs, and a no-slip regime for rough bases ($R_a \gtrsim 0.6$) where the basal layer is not easily visible but deviations from Bagnold's profile still exist. These flow regimes are roughly consistent with the slip-to-no-slip transition found in \citet{Jing2016} and the threshold values of $R_a$ are insensitive to flow conditions (slope angle and flow thickness).

The kinematics profiles (velocity, shear rate and temperature) are examined. When slip occurs, the bulk velocity profile (which follows Bagnold's scaling) is shifted by an effective slip velocity, and the basal-layer velocity profile appears to be exponential. This feature corresponds to an exponential increase in the basal-layer shear rate and granular temperature profiles. Based on these observations, we propose a generalized Bagnold-like velocity profile considering the slip velocity and the characteristic thickness of the basal layer. Furthermore, a Navier-like slip law is developed for all roughness and flow conditions, which can be viewed as an extension of a previous slip law for flat-base inclined flows \citep{Artoni2012}. We also examine various granular temperature-based slip laws \citep{artoni_effective_2015,breard_basal_2020,gollin_extended_2017}, showing that a slip boundary condition in the granular kinetic theory \citep{gollin_extended_2017} can capture the overall trend of the results.

Rheological data (stress, friction, and volume fraction) extracted from the basal layer are analysed in various frameworks, including the local $\mu(I)$ rheology, Bagnold's stress scaling, and kinetic theory-based stress models. It is found that local rheology breaks down in the dilute, collision-dominated basal layer, while kinetic theory models can well describe the pressure and shear stress as functions of shear rate, temperature, and volume fraction. These results indicate that the balance of fluctuation kinetic energy controls the basal layer dynamics and invite future development of kinetic theory models to fully resolve the kinematics and rheological profiles of the basal layer of granular flows. 

Finally, while we focus on dry granular flows consisting of spherical particles down a rigid substrate, more realistic or complicated situations involving the effects of particle shape, segregation, interstitial fluids and erodible substrates can be considered in the future. Dynamics of thin-layer granular flows on smooth and rough planes, where the boundary layer effects dominate over the bulk flow, along with its implications for seismic signal generation in geophysical flows~\citep{arran_simulated_2024}, is also an interesting future direction.

\backsection[Acknowledgements]{We would like to thank Yifei Duan, Gengchao Yang, Thomas Pähtz and Diego Berzi for helpful discussions. T. Wang acknowledges the financial support received as a visiting student at Tsinghua Shenzhen International Graduate School, Tsinghua University. The numerical simulations were conducted using the high-performance computer clusters (HPC) at The University of Hong Kong. 
}

\backsection[Funding]{Supported by the National Natural Science Foundation of China (12472412) and the Research Grants Council of Hong Kong (17205821, 17205222, 17200724).}

\backsection[Declaration of interests]{The authors report no conflict of interest.}


\backsection[Data availability statement]{The data that support the findings of this study, as well as the open-source codes for reproducing the simulations, are available from the corresponding author upon reasonable request.}

\backsection[Author ORCIDs]{\\T. Wang, https://orcid.org/0000-0001-9427-8609; \\L. Jing, https://orcid.org/0000-0002-1876-1110; \\C.Y. Kwok, https://orcid.org/0000-0002-1019-7551; \\Y.D. Sobral, https://orcid.org/0000-0001-5775-1292; \\T. Weinhart, https://orcid.org/0000-0002-2248-7644; \\A.R. Thornton, https://orcid.org/0000-0001-8199-7014.}


\appendix

\section{Results of same $R_a$ but different base particle arrangements} \label{app:base}

\begin{figure}
  \centerline{\includegraphics[width=1\linewidth]{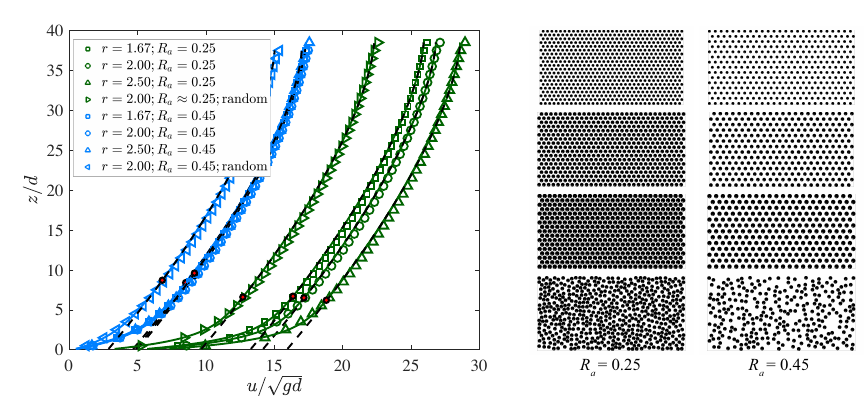}}
  \caption{Velocity profiles for the same $R_a$ but different base particle arrangements ($H/d=40, \theta=23^\circ$). Detailed base configuration parameters are listed in Table~\ref{tab:configuration}.}
\label{fig:sameRa}
\end{figure}

The base roughness for granular flow is difficult to define, as it depends on the flow-to-base-particle size ratio $r$, base particle spacing $\epsilon$, and the specific spatial arrangement (regular or random) of the base particles. To demonstrate that $R_a$ is a good indicator of the base roughness, figure~\ref{fig:sameRa} presents results for the same $R_a$ but different combinations of $r$ and $\epsilon$ with a regular or random arrangement ($H/d=40, \theta=23^\circ$). It shows that, although the geometric features of the base vary significantly (see right panels), the velocity profiles are nearly identical for $R_a=0.45$ (slip-to-no-slip transition regime) and are reasonably close for $R_a=0.25$ (slip regime). Only the randomly generated bases (where the void areas have a distribution, see \S\ref{sec:method-Ra}) seem to always induce smaller basal slip, perhaps because larger voids have a stronger influence on the slip velocity. Nevertheless, the difference in terms of the slip velocity is within the range of uncertainties (see figure~\ref{fig:slip}d of the main text). Therefore, although an alternative roughness indicator that better collapses the data for $R_a=0.25$ warrants further studies, the use of $R_a$ effectively unifies the influence of $r$ and $\epsilon$, which are major parameters determining the base roughness \citep{Jing2016}. 

\section{Modified Bagnold's profiles considering basal slip velocity}\label{app:Bagnold}

Here we present the derivation of Bagnold's velocity profile with slip velocity, as well as the shear rate and temperature profiles, assuming that the entire flow obeys the $\mu(I)$ rheology. An alternative form of Bagnold's velocity profile using the rheology $\tau\propto\rho_pd^2\dot\gamma^2$ leads to the same basic scaling as below but different fitting constants~\citep{Silbert2001}.

For steady, fully developed flows down an incline of slope $\theta$ in the $xz$-plane, integrating the momentum balance equations along the $z$-direction leads to $\tau=\phi\rho_pg\sin\theta(h-z)$ and $P=\phi\rho_pg\cos\theta(h-z)$, such that $\mu=\tau/P=\tan\theta$ is constant. According to the $\mu(I)$ local rheology~\citep{forterre_flows_2008}, where $\mu$ and $\phi$ are functions of $I$, the volume fraction $\phi$ and inertial number $I$ are also constant along $z$. 

By identifying that $I=\dot\gamma d/\sqrt{P/\rho_p}$, the shear rate $\dot\gamma$ can be expressed as
\begin{equation}
    \dot\gamma := \frac{\partial u}{\partial z} = \frac{I\sqrt{P/\rho_p}}{d} = \frac{I\sqrt{\phi g\cos\theta}}{d} \left(h-z\right)^{1/2}
  \label{eq:gamma}
\end{equation}

Assuming a basal slip velocity, $u(z=0)=u_b$, and integrating (\ref{eq:gamma}) along $z$, we have
\begin{equation}
    u = u_b+\frac{2I\sqrt{\phi g\cos\theta}}{3d}\left[h^{3/2} - (h-z)^{3/2}\right]
\end{equation}

Substituting the empirical correlation $\sqrt{T}=\alpha\dot\gamma d$ for steady granular flows into (\ref{eq:gamma}), the granular temperature profile can be expressed as
\begin{equation}
    T = \alpha^2I^2\phi\rho_p g\cos\theta(h-z)
\end{equation}

In the main text (\S\ref{sec:kinematics}), $\phi$ and $I$ are obtained as constants from the bulk, $u_b$ is obtained as a fitting parameter of the velocity profile, and $\alpha$ is obtained by fitting the $\sqrt{T}$ vs.\ $\dot\gamma d$ results. 

\section{Derivation of Bagnold's stress from kinetic theory}\label{app:stress}

\citet{Bagnold1954} studied steady, homogeneous sheared granular flows and proposed the stress scaling $\tau\propto P\propto\rho_pd^2\dot\gamma^2$, where the proportionality coefficients are unknown functions of $\phi$. Recently, \citet{Berzi2024} recovered this scaling from an extended kinetic theory for granular gases of frictional particles, where the constitutive relations for the stresses are
\begin{equation}
    p=\rho_p\phi[1+2(1+e_n)\phi g_0]T,
    \label{p KTGF}
\end{equation}
\begin{equation}
    \tau=\rho_p\frac{8J\phi^2g_0}{5\pi^{1/2}}dT^{1/2}\dot{\gamma},
    \label{tau KTGF}
\end{equation}
where
\begin{equation}
    J={\left\{\frac{1+e_n}{2}+\frac{\pi}{32}\frac{[5+2(1+e_n)(3e_n-1)\phi g_0][5+4(1+e_n)\phi g_0]}{[24-6(1-e_n)^2-5(1-e_n^2)\phi^2g_0^2]}\right\}\left[1+\frac{\pi}{12}\frac{5-3(2-e_n)}{3-e_n}\right]}
\end{equation}
is a function containing the radial distribution function $g_0$,
\begin{equation}
    g_0=\left[ 1-\mathcal{H}(\phi-0.4)\left(\frac{\phi-0.4}{\phi_c-0.4}\right)^2\right]\frac{2-\phi}{2(1-\phi)^3}+\mathcal{H}(\phi-0.4)\left(\frac{\phi-0.4}{\phi_c-0.4}\right)^2\frac{2}{\phi_c-\phi},
    \label{g_0}
\end{equation}
where $\mathcal{H}(\cdot)$ is the Heaviside function and $\phi_c$ is a critical volume fraction, which is generally a funciton of particle friction $\mu_p$. We choose $\phi_c=0.58$ according to our input ($\mu_p=0.5$) and the empirical relations provided by \citet{Berzi2024}.

Now consider the balance equation for the fluctuation kinetic energy in a unidirectional flow in the $xz$-plane,
\begin{equation}
    \frac{3}{2}\rho_p\phi\frac{\mathrm{d}T}{\mathrm{d}t} = \tau\dot{\gamma} -\frac{\partial{Q}}{\partial{z}} - \Gamma,
    \label{energy conservation}
\end{equation}
where $-\partial{Q}/\partial{z}$ refers to the diffusion of energy through particle agitation, $\tau\dot{\gamma}$ represents the production of energy through the work of shear stress and $\Gamma$ is the rate of energy dissipation. In steady ($\textrm{d}T/\textrm{d}t=0$) and homogeneous ($\partial{Q}/\partial{z}=0$) flows, this equation reduces to
\begin{equation}
\tau\dot{\gamma} - \Gamma=0,
    \label{energy balance}
\end{equation}
which states that the fluctuation energy generated due to shear is dissipated through particle interactions. Although the granular temperature is clearly inhomogeneous in the basal layers (\S\ref{sec:kinematics}), we carry on with the assumption of $\partial{Q}/\partial{z}=0$ for simplicity and, as shown in the main text, the model predictions are fairly good. More sophisticated analyses without this assumption can be found in~\citet{kumaran_dense_2008} and \citet{gollin_extended_2017}.

The rate of dissipation, $\Gamma$, for inelastic, frictional spheres is expressed as
\begin{equation}
    \Gamma=\rho_p\frac{12(1-e_\textrm{eff}^2)\phi^2g_0}{\pi^{1/2}}\frac{T^{3/2}}{L},
    \label{energy dissipation}
\end{equation}
where $e_\textrm{eff}=e_n-(\pi/2)\mu_p+(9/2)\mu_p^2$ is the effective coefficient of restitution for nearly elastic spheres and $L$ is the correlation length that accounts for correlated particle velocity fluctuations at $\phi>0.49$,
\begin{equation}
\frac{L}{d}=\left[\frac{2J}{15(1-e_\textrm{eff}^2)}\right]^{1/2}\left[1+\frac{26(1-e_\textrm{eff})}{15}\frac{\mathcal{H}(\phi-0.49)(\phi-0.49)}{0.64-\phi}\right]^{3/2}\frac{d\dot{\gamma}}{T^{1/2}},
  \label{correlation length}
\end{equation}
where $0.64$ is the value of $\phi$ at the random close packing. 

Substituting \eqref{tau KTGF}, \eqref{energy dissipation} and \eqref{correlation length} into \eqref{energy balance} yields an expression for $T$,
\begin{equation}
    T=\frac{2J}{15(1-e_\textrm{eff}^2)}\left[1+\frac{26(1-e_\textrm{eff})}{15}\frac{\mathcal{H}(\phi-0.49)(\phi-0.49)}{0.64-\phi}\right]d^2\dot{\gamma}^2,
    \label{T gamma}
\end{equation}
which somewhat explains the observed $\sqrt{T}\propto\dot\gamma d$ correlation in \S\ref{sec:kinematics}. Finally, substituting \eqref{T gamma} into \eqref{p KTGF} and \eqref{tau KTGF} recovers Bagnold's stress models,
\begin{equation}
    p=\frac{2J\phi[1+2(1+e_n)\phi g_0]}{15(1-e_\textrm{eff}^2)}\left[1+\frac{26(1-e_\textrm{eff})}{15}\frac{\mathcal{H}(\phi-0.49)(\phi-0.49)}{0.64-\phi}\right]\rho_pd^2\dot{\gamma}^2
  \label{p gamma}
\end{equation}

\begin{equation}
    \tau=\frac{8J\phi^2g_0}{5\pi^{1/2}}\left[\frac{2J}{15(1-e_\textrm{eff}^2)}\right]^{1/2}\left[1+\frac{26(1-e_\textrm{eff})}{15}\frac{\mathcal{H}(\phi-0.49)(\phi-0.49)}{0.64-\phi}\right]\rho_pd^2\dot{\gamma}^2
  \label{tau gamma}
\end{equation}

The expressions \eqref{p KTGF}, \eqref{tau KTGF}, \eqref{p gamma} and \eqref{tau gamma} are repeated in \S\ref{sec:rheology} and used in figure~\ref{fig:rheology} to predict stresses in both the bulk and basal layers. Moreover, it is possible to derive a $\phi$-$I$ relationship using \eqref{p gamma} noting that the inverse square of $I$ is precisely $p/(\rho_p d^2\dot{\gamma}^2)$; this relationship is also included in figure~\ref{fig:rheology}(b) for comparison.

\end{document}